\documentclass[preprint,showpacs,preprintnumbers,amsmath,amssymb]{revtex4}

\usepackage[latin1]{inputenc}
\usepackage{graphicx}
\usepackage{dcolumn}
\usepackage{bm}

\begin{document}

\title{Exact expressions of mean first-passage times and splitting probabilities
for random walks in bounded rectangular domains}

\author{S. Condamin}
\author{O. B{\' e}nichou}
\affiliation{Université Pierre et Marie Curie-Paris6,
Laboratoire de Physique Théorique de la Matière Condensée, 
UMR-CNRS 7600, case courrier 121, 4 Place Jussieu, F-75005
Paris}

\date{\today}


\maketitle

Recently, we have proposed a novel computation method
 of first-passage times of a random walker between a starting site and a target
 site of regular bounded lattices of arbitrary shape\cite{PRL}. Such first 
passage time properties are for example crucial to describe the kinetics of 
diffusion limited reactions in confined media\cite{Rice,Redner}. The obtained expressions 
involve pseudo-Green functions\cite{Barton}, which have been estimated according to 
different approximation schemes.
In this note, we give the exact expression of these pseudo-Green functions in 
both cases of
a rectangular domain with reflecting boundaries and a rectangular domain with 
periodic boundary conditions. This allows us to provide exact and explicit 
expressions
of mean first-passage times with one or two target sites.

 Consider first a random walker starting
at the source $S$, of position ${\bf r}_S$ of a rectangular regular lattice with reflecting boundary conditions.
The mean 
time it takes to reach the target $T$, of position ${\bf r}_T$, for the first
time, is given by\cite{PRL} 
\begin{equation}
\langle \mathbf{T}\rangle = N [ H({\bf r}_T|{\bf r}_T) - H({\bf
r}_T|{\bf r}_S)  ]
\end{equation} 
where $H$ is the pseudo-Green function, which satisfies
\begin{equation}
H({\bf r}_i|{\bf r}_j) = \frac1\sigma \sum_{\langle k,i\rangle} 
H({\bf r}_k|{\bf r}_j)+ 
\delta_{ij} - \frac1N,
\label{pseudogreen}
\end{equation}
$\sigma$ being the coordination number of the lattice, namely 4 for a 2-D 
lattice,
or 6 for a 3-D lattice. In this expression, the sum runs over all the 
neighbours $k$ of the site $i$, 
considering that a site which is at the boundary of the domain is its own
neighbour. $N$ is the total number of sites of the lattice. 
In the presence of two absorbing 
targets, the mean time it takes to reach either of the two targets is:
\begin{equation}
\label{time}
\langle \mathbf{T}\rangle = N 
\frac{(H_{01}-H_{1s})(H_{02}-H_{2s}) - (H_{12}-H_{2s})(H_{12}-H_{1s})}
{H_{01}+H_{02}-2H_{12}}
\end{equation}
while 
the eventual hitting probabilities $P_i$ to reach the target $i$ writes
\begin{equation}
\left\{
\begin{array}{l}
P_1 = \frac{H_{1s}+H_{02}-H_{2s}-H_{12}}{H_{01}+H_{02}-2H_{12}} \\
P_2 = \frac{H_{2s}+H_{01}-H_{1s}-H_{12}}{H_{01}+H_{02}-2H_{12}} \\
\end{array}
\right.
\end{equation}
where  $H_{12} = H({\bf r}_{T_1}|{\bf r}_{T_2})$ and, 
for $i = 1$ or $2$,  $H_{is} = H({\bf r}_{T_i}|{\bf r}_S)$, 
$H_{0i} = H({\bf r}_{T_i}|{\bf r}_{T_i})$.

The exact expression  of the pseudo-Green function $H$
involved in previous equations 
may be computed explicitely with the help of Fourier analysis. 
For a 2D domain with $X$ sites in the $x$ direction and $Y$ sites in the $y$  
direction, $H$ writes
\begin{eqnarray}
H({\bf r}|{\bf r}')&=& \frac1N \sum_{m=1}^{X-1} \sum_{n=1}^{Y-1}
\frac{4 \cos\frac{m\pi x'}{X}\cos\frac{n\pi y'}{Y}
\cos\frac{m\pi x}{X}\cos\frac{n\pi y}{Y}}{
1-\frac12\left(\cos\frac{m\pi}{X}+\cos\frac{n\pi}{Y}\right)}\nonumber \\
&& + \frac1N \sum_{m=1}^{X-1}\frac{4\cos\frac{m\pi x'}{X}\cos\frac{m\pi x}{X}}{
1 - \cos\frac{m\pi}{X}} \\
&& + \frac1N \sum_{n=1}^{Y-1}\frac{4\cos\frac{n\pi y'}{Y}\cos\frac{n\pi y}{Y}}{
1 - \cos\frac{n\pi}{Y}} \nonumber 
\end{eqnarray}
where $x$ and $y$ are the coordinates of ${\bf r}$ and  $x'$ and $y'$ those  
of ${\bf r'}$. 
Here, the coordinates
of the left-bottom corner are taken equal to $(1/2,1/2)$, so that 
all the coordinates are half-integers. 

\begin{figure}[t]
\centering\includegraphics[width = .7\linewidth,clip]{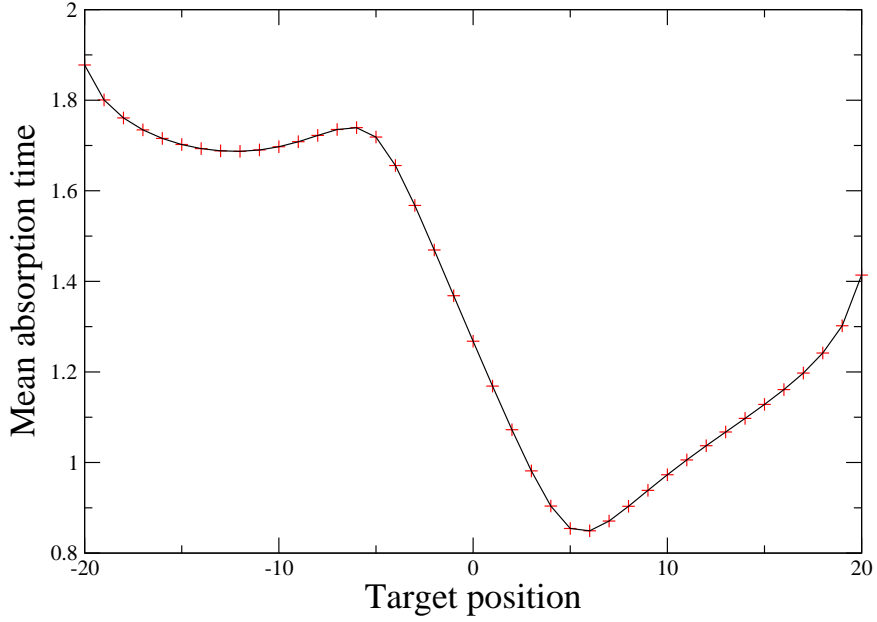}
\caption{2D Two-target simulations (red crosses) 
vs. theory 
(plain line). One target is fixed at (-5,0); The source is fixed at (5,0); 
The other target is at (x,3). The domain is a square of
side 41, the middle is the point (0,0). 
The absorption time is normalized by the number of sites $N$.}
\label{simu}
\end{figure}

Figure \ref{simu} shows the mean absorption time $\langle\mathbf{T}\rangle$
given by eq. (\ref{time}) as a function of the target
positions together with numerical simulations
in the case of a 2D square domain.
These simulations have been performed with a method based on the exact 
enumeration method\cite{Majid}.

Similar results can be obtained in the case of a rectangular domain with 
periodic boundary conditions.
Here,  the pseudo-Green function is
\begin{equation}
H({\bf r}|{\bf r}') = \frac1N \sum_{m=0}^{X-1}\sum_{n=\delta_{0m}}^{Y-1}
\frac{\exp\frac{2im\pi(x-x')}{X}\exp\frac{2ni\pi(y-y')}{Y}}{1-\frac12\left(
\cos\frac{2m\pi}{X}+\cos\frac{2n\pi}{Y}\right)}
\end{equation}
where $\delta_{0m}$ is the Kronecker delta. 

Note that our results can be easily extended to the 3D case. 

We gratefully thank S. Redner for suggesting us the simulation method.

\end{document}